# A multi-objective optimization framework for terrain modification based on a combined hydrological and earthwork cost-benefit

Hanwen Xu[1*], Mark Randall[1], Lei Li[2], Yuyi Tan[3], Thomas Balstrøm[1]

1. Department of Geosciences and Natural Resources Management, University of Copenhagen, Frederiksberg, Denmark
2. Department of Computer Science, University of Copenhagen, Copenhagen, Denmark
3. Amazon.com Services LLC, Seattle, USA

**Abstract:** The escalating risk of urban inundation has drawn increased attention to urban stormwater management. This study proposes a multi-objective optimization for terrain modification, combining the Non-dominated Sorting Genetic Algorithm II (NSGA-II) with digital elevation model (DEM)-based hydrological cost factor analysis. To reduce the precipitation's erosive forces and runoff's kinetic energy, the resulting framework offers the possibility of efficiently searching numerous solutions for trade-off sets that meet three conflicting objectives: minimizing maximum flow velocity, maximizing runoff path length and minimizing earthwork costs. Our application case study in Høje Taastrup, Denmark, demonstrates the ability of the optimization framework to iteratively generate diversified modification scenarios, which form the reference for topography planning. The three individual objective preferred solutions, a balanced solution, and twenty solutions under regular ordering are selected and visualized to determine the limits of the optimization and the cost-effectiveness tendency. Integrating genetic algorithms with DEM-based hydrological analysis demonstrates the potential to consider more complicated hydrological benefit objectives with open-ended characteristics. It provides a novel and efficient way to optimize topographic characteristics for improving holistic stormwater management strategies.

# 1. Introduction

Effective urban stormwater management (USM) is crucial in the context of climate change and urbanization to reduce urban inundation risks and mitigate the ongoing degradation of aquatic ecosystems (Chen et al., 2016; Fletcher et al., 2013; Lee and Bang, 2000; Miltner et al., 2004; Yin et al., 2021). The topographic characteristics drastically impact the hydrological process, including confluence, retention and infiltration through changes in elevation, slope or aspect, and so on, which further influences USM planning and design implementation (Eckart et al., 2018; Lee et al., 2012). Topographic factors are also part of the most common yet challenging parts of landscape design and urban planning (Burns et al., 2012). Therefore, understanding and modifying the topography of urban surfaces is a critical component in developing effective strategies for USM.

As the surface of urbanization progresses, the surface elevation frequently changes due to terrain modification activities, leading to related hydrological performance and benefits. At the macro-scale, the topography affects the delineation of regional catchments and the determination of flooding risk areas. At the micro-scale, the terrain design influences the specific runoff flow paths and the spatial distribution of sinks or water bodies. Thus, rational terrain planning and modification could positively influence hydrological processes, thereby reducing the adverse impacts of extreme downpours and flooding events. Nowadays digital elevation models (DEMs), as available and accurate geographic data, show detailed raster information on elevations for analyzing, interpreting, and optimizing urban topography. Although raster-based DEM analysis leads to a representation of surface flows in a 2D network not involving any hydrodynamic components, it provides a concise and visual overview of drainage basins, the location of sinks and accumulated downstream flow within a landscape (Balstrøm and Crawford, 2018).

Simulation and optimization of hydrological benefits from terrain modifications is a widely explored research topic (Chen et al., 2021; Fletcher et al., 2013; Salvan et al., 2016). Numerous studies have focused on hydrological and hydraulic performance optimization using platforms like the Storm Water Management Model (SWMM), Soil and Water Assessment Tool (SWAT) and MIKE URBAN (Chen et al., 2018; Randall et al., 2020; Sidek et al., 2021). Studies determine the mix of solutions to maximize hydrological performance and cost-benefits as much as possible optimizing the combination of complex factors (hydrological or non-hydrological) to improve the comprehensive influence of stormwater management (Eckart et al., 2018; Johnson and Geisendorf, 2019; Liu et al., 2023b). Such increased complexity research often turns into multi-objective optimization (MOO) problems in which the objectives are nonlinear and inter-constrained, and often complicated to the point of intractability (Nesshöver et al., 2017; Xu et al., 2023; Zhang and Chui, 2018). Liu et al. (2023b), Wang et al. (2023), and Yao et al. (2022) conducted studies related to the optimization of green-gray infrastructure around the hydrological, capital and ecological objectives influenced by the layout of these infrastructures. Saadatpour et al. (2020), Sun et al. (2022), Xu et al. (2017), and Yang et al. (2023) addressed the MOO between the spatial layout and multiple benefits of stormwater management facilities. However, there has been limited research integrating the analysis of terrain data with hydrological benefits in different scales by establishing MOO frameworks. This research gap can be attributed to factors, including 1) There is a lack of corresponding quantitative hydrological evaluation for the terrain modification process; 2) It is an issue for terrain designers (e.g., landscape architects) accustomed to empirical observation to handle early terrain design with hydrological calculations and assess all possibilities by any enumeration method (Chen et al., 2016). No matter what the reason is, the terrain optimization problem is a challenge not only because of the exponentially large number of variables involved but also the increasing number of objectives (Cao et al., 2011). To overcome these, integrating MOO algorithms into the terrain modification process shows potential.

Terrain modification optimization is complicated by the key point that it involves not only where to allocate modification activities but also how much earthwork volume to fill or cut in allocated locations. If the investment is disregarded when pursuing ideal hydrological outcomes, it is always possible to identify more optimized modification solutions. Therefore, the question is how to achieve favorable hydrological benefits (e.g., runoff paths and volumes, flow velocities and sink distributions) while identifying modification locations and minimizing total modification operations which result in reduced costs. As one kind of MOO algorithm, the non-dominated sorting genetic algorithm II (NSGA-II) is effective for its flexibility and adaptability in addressing this type of MOO problem. While numerous studies have employed genetic algorithms for optimizing hydrological benefits against costs, research specifically utilizing genetic algorithms to optimize topographic features is limited (Leng et al., 2021; Tang et al., 2022; Xie et al., 2022; Xu et al., 2018). Consequently, this study aims to address the challenges and limitations related to the use of MOO in determining hydrological-cost benefits. Specifically, these are to:

1. Develop the optimization framework that integrates genetic algorithms with DEM-based hydro-morphometric analysis, while considering various cost factors. It aims to explore the feasibility, efficiency and reliability of the obtained solution sets.

2. Quantify the comprehensive impacts of this terrain modification optimization framework addressing the uncertainties and time-consuming repetitive nature of the terrain modification.

To achieve these goals, this research combines the NSGA-II algorithm with hydro-morphometric analysis based on DEM data in an integrated Python platform. Moreover, through an efficient optimization process, this research aims to assist in design decision-making by rapidly generating and visualizing potential solutions to be replicated within corresponding zones. Thus, this approach holds the potential to improve the implementation of USM strategies by incorporating optimized terrain features into the urban landscape.

## 2. Methodology

### 2.1. Study area

The study area is 39 ha located in Høje Taastrup municipality in the western suburbs of Copenhagen, Denmark (55°36'12.75''N-55°36'35.66''N, 12°11'30.55''E-12°12'17.28''E) (Figure 1). The site experiences a temperate maritime climate, typical of the Copenhagen area. This area was chosen as a study site due to: 1) it is in the upstream area of the Region Hovedstaden in terms of macro-scale watershed location, so there is no upstream external flow contribution to the site under rainfall events; 2) the surface is predominantly green fields with a few farmhouses and connection roads so that to be approximatively treated as a uniform green space underlying surface throughout the analysis and optimization procedures; 3) the site is dominated by clay-rich top- and subsoils low in hydraulic conductivity susceptible to slow water infiltration rates and, thus, high runoff volumes in stormwater situations.

The DEM used for the study was resampled from a cell size of 0.4 m to a resolution of 10 m * 10 m to manage computational requirements and maintain a balance between calculation accuracy and efficiency. The terrain has gentle slopes and elevation values ranging from 35.04 to 48.19 m which are generally low in the southeast and high in the northwest. Overall, the study area's geographical characteristics serve as the foundation for the optimization framework, offering insights into the terrain, climate and land cover.

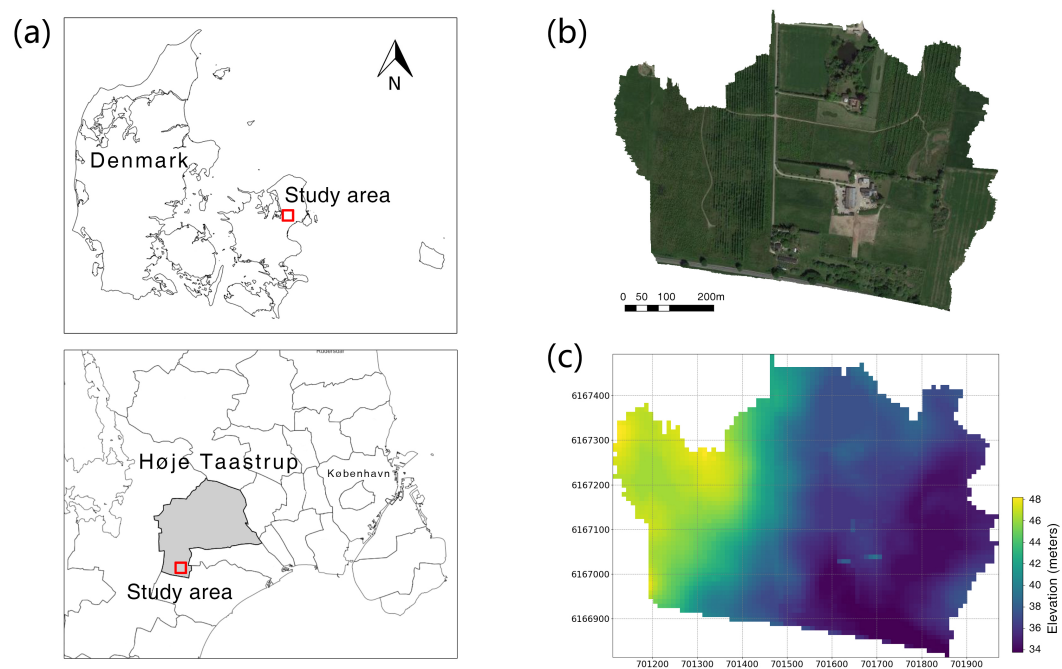

Figure 1 (a) Location of the study area; (b) study area orthophoto © Danish Agency for Data Supply and Infrastructure; (c) DEM data © Danish Agency for Data Supply and Infrastructure.

## 2.2. Optimization framework

Figure 2 depicts the workflow of the terrain modification multi-objective optimization, refers to as TMMOO. In the first step, the DEM data information is loaded into the Python integrated development environment (IDE), and the initialized scenario is analyzed and visualized. In the second step, the TMMOO module is assigned variables, related constraints, and objective functions. In the third step, a NSGA-II-based iteration search for the TMMOO solutions is set dealing with the decision-making and visualization processes. The components are explained in detail in sub-sections 2.2.1-3.

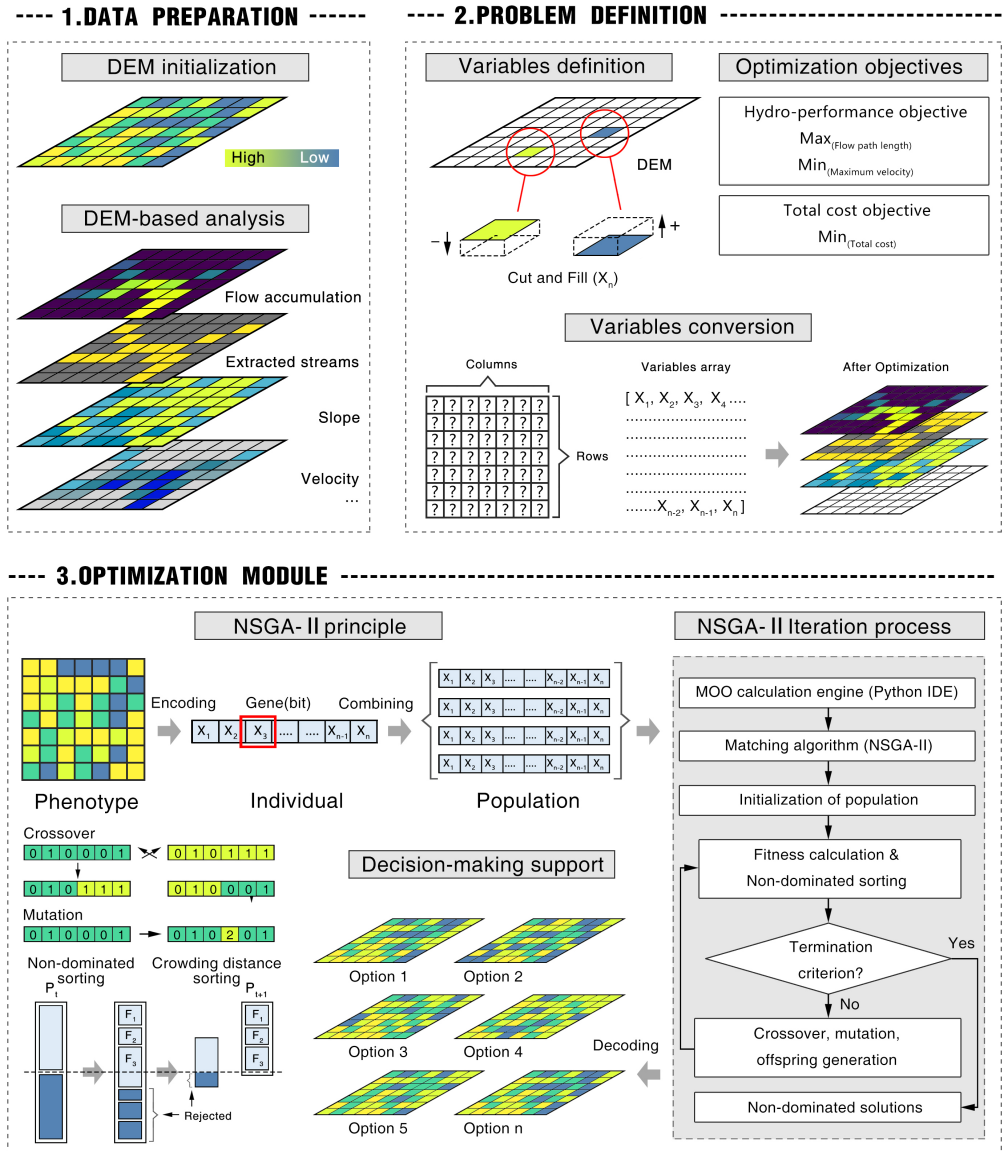

Figure 2 Flowchart of TMMOO framework and procedures

### 2.2.1. Objective functions

Terrain modification optimization needs to meet many different objectives based on a comprehensive understanding of the requirements pertaining to terrain and spatial planning. The terrain modification which we focus on here to achieve better hydrological benefits by changing grid cell elevations, involves evaluation of three objective functions (OF): 1) maximizing flow path length; 2) minimizing maximum runoff velocity; 3) minimizing earthwork cost (Eq1).

$$\begin{cases} Max(Flow\ path\ length) \\ Min(Maximum\ runoff\ velocity) \\ Min(Earthwork\ costs) \end{cases}$$

Maximizing the flow path length associated with terrain modification is one of the objectives. The flow path is defined by flow accumulation analysis and threshold delineation (Ariza-Villaverde et al., 2015). An increased flow path facilitates rainwater retention, infiltration, and storage at the surface by prolonging the residence time of stormwater. Also, it facilitates water quality improvement through the increased potential for stormwater treatment and filtration processes.

Another objective is minimizing the maximum runoff velocity. Except for the surface roughness which is usually decided by natural conditions, velocity tends to correlate with slope gradient and length. Higher slopes and lengths always produce a faster velocity. Avoiding excessive runoff velocities during storm events offers notable advantages, including decreasing negative effects such as surface scouring and soil erosion, thus protecting soil fertility and minimizing sediment transport into water bodies (Nicola et al., 1999; Yang et al., 2023).

Minimizing cost is a typical objective function found in MOO frameworks, which is also applied in this study (Eckart et al., 2018; Shishegar et al., 2018). Cost management often poses a potential mutual exclusion with other objectives since more investments usually lead to better returns. Finding trade-off cost-benefit solutions is tricky for investors or any stakeholders. The calculation methods of OFs are as follows:

1) OF1: maximizing flow path length

The initial DEM that contains the terrain elevation in each grid cell is used as an input to identify the surface flow. A flow direction based on the so-called D8 method, first introduced by (O'Callaghan and Mark, 1984), is assigned to each grid cell depending on the steepest downhill slope direction derived from the DEM using a 3x3 moving window after corrections of the DEM's inaccuracies and anomalies (Figure 3). Based on flow direction, a flow accumulation figure is obtained, and cells whose flow accumulation values are higher than or equal to a defined threshold value will comprise the flow path (Figure 4). By referencing past studies, we chose 2% of the maximum flow accumulation value as the threshold value (Ariza-Villaverde et al., 2013, 2015; Dávila-Hernández et al., 2022) which is 14.36 for the study area. Finally, the flow path length is quantified by counting the number of grid cells higher than the threshold.

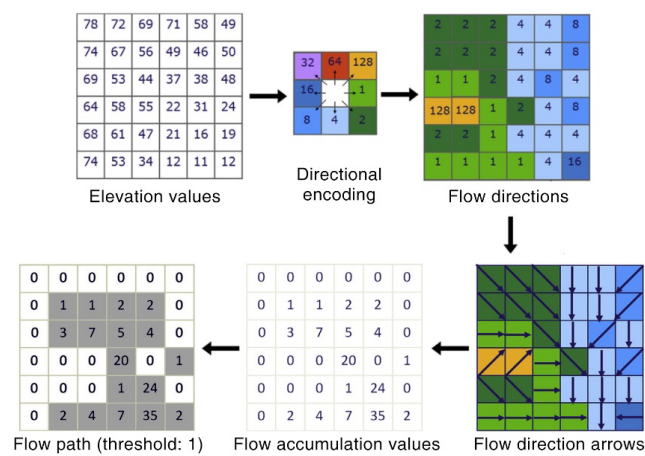

Figure 3 The process of deriving flow paths and flow accumulation values from an elevation raster (adopted from Esri, 2017)

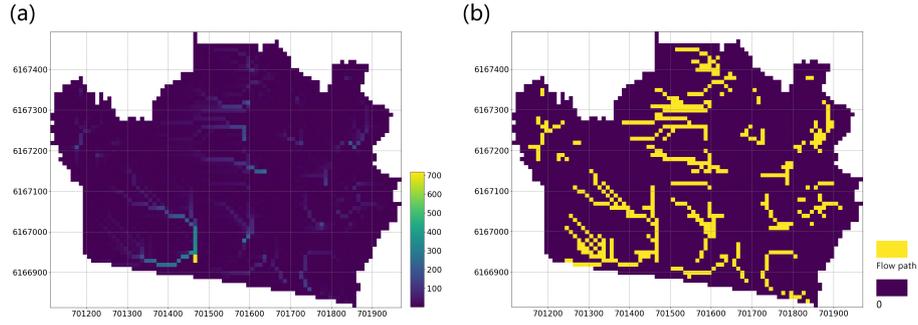

Figure 4 Flow accumulation analysis (a) and flow path (b)

2) OF2: Minimizing maximum runoff velocity

The runoff velocity is calculated based on slope, D8 flow accumulation and the steady-state continuity equation, incorporating Manning's coefficient (Figure 5) (Eq2) (Melesse and Graham, 2004). Then the maximum value is found (maximum runoff velocity) by screening over the data.

$$V_i = \left[\frac{S^{\frac{1}{2}}}{n}\left(\frac{Q}{B}\right)^{\frac{2}{3}}\right]^{\frac{3}{5}}$$

Where $V_i$ is the flow velocity (m/s); $Q_i$ is the cumulative discharge (m³/s) through the cell, obtained by multiplying upstream flow contributions and the precipitation intensity in unit area for that cell; $S_i$ is the local slope value (%); n is the local Manning coefficient; B is the channel area (m²). It was assumed that channels had a rectangular cross-section with depth and effective width obtained from parameter sensitivity analysis reference (Melesse and Graham, 2004).

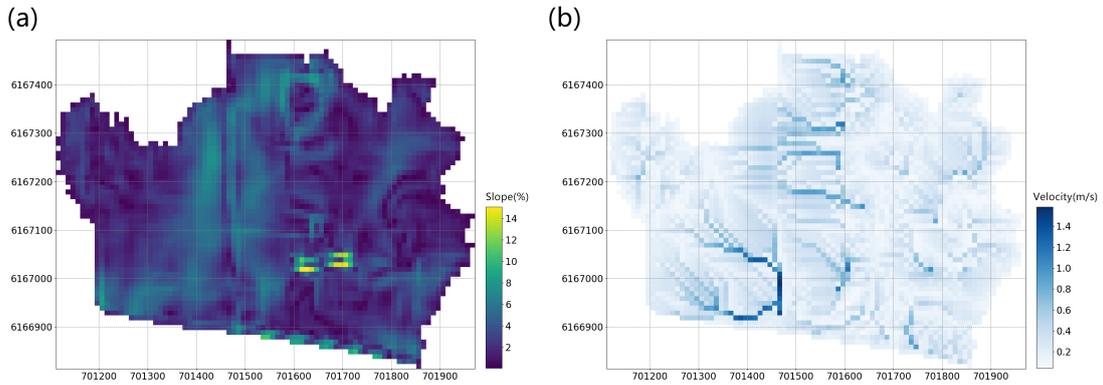

Figure 5 Slope (a) and runoff velocity (b)

3) OF3: minimizing earthwork costs:

We used a very straightforward and simple method of calculating costs. The modification's total cost is determined by the net earthwork volume and unit price (Eq3).

$$C_{total} = \sum_{i=1}^{N_{var}} |E_i| \times R \times P_i$$

Where $C_{total}$ is the total cost of terrain modification (DKK); $E_i$ is the elevation change value (m), R is a cell's area, 100m²; $P_i$ is the unit price of net earth movement (DKK/m³).

### 2.2.2. Variables and constraints

The main task of this optimization framework is to determine the modification volume and location of the site, so the variables are the change values assigned to each grid cell in the DEM. By modifying the DEM's elevation values, each variable represents the magnitude of the terrain elevation's physical change. The number of variables ($N_{var}$) is equal to the number of grid cells in the DEM, calculated as the product of the number of rows ($N_{row}$) and columns ($N_{col}$) for a rectangular area (Eq4). If the study area is not rectangular, it needs to subtract the number of grid cells with no-data elevation values ($N_{non\text{-}data}$). In our study, the dataset covers a total of 3903 valid raster cells. The precision of the terrain modification is determined by both the data resolution and the cell number. A higher resolution DEM indicates a finer result of detail but also increases the $N_{var}$, potentially affecting the optimization efficiency.

$$N_{var} = N_{row} \times N_{col} - N_{non-data}$$

The constraints involved in the optimization framework have two categories: variable constraints and objective constraints. Variable constraints define the range for each variable, in this study, the change volume for each elevation cell is between cut 2 meters and fill 2 meters. The modified DEM is computed by adding the changed value to the original DEM. The positive values correspond to elevation increases, representing earthwork filling, while negative indicate decreases representing earthwork cutting.

### 2.2.3. Optimization algorithm and platform

As mentioned previously, the NSGA-II, a widely adopted algorithm that effectively addresses MOO problems is chosen for this optimization study (Deb and Jain, 2014; Deb et al., 2002). The main loop process includes several parts. First, it randomly generates an initial population of size n. Secondly, assess each individual and sort solutions by fitness calculation and non-dominated sorting. Crossover and mutation based on the parent population apply new candidates for the offspring population. As previous and current population members are included, the new offspring population is generated based on the non-dominance relationship and crowding distance sorting (Cao et al., 2011; Deb et al., 2002). Then, solutions from the generation are chosen next followed by solutions from the generation, and so on. The optimization scenarios under setting objectives are generated until iteration stops or end conditions are achieved (Figure 2). In this process, the population size, crossover and mutation probability, and iteration number are conclusive for the TMMOO performance and computational speed.

The Pymoo library (Blank and Deb, 2020) is employed to provide a built-in implementation of the NSGA-II algorithm. Pymoo facilitates the incorporation of the defined variables, constraints, and objectives into the optimization process. Additionally, the Whitebox workflow (WbW) (Lindsay, 2016) is utilized, with a topographic and hydro-morphometric analysis module to serve as support for the OFs. WbW is a Python library for advanced geoprocessing, including functions for GIS and remote sensing analysis and for manipulating geospatial data (raster, vector and LiDAR). We integrated these tools into PyCharm platforms to allow for an efficient and comprehensive optimization process.

## 2.3. Solution process

The solution process outlines the procedure employed to identify the optimal solutions within the optimization module (Figure 2). The process consists of NSGA-II algorithm matching, initialization of population, optimization parameter selection and loop setting. In this process, the population size is 200 and the offspring size is 100. The initialized population represents 200 random DEMs of the same size as the study area, and then offspring are generated at a time to be screened along with the parents. This elite iteration leads to a progressive search for better solutions. We set the iteration number to 300 and chose simulated binary crossover (crossover probability = 0.9, eta =15) by referencing past studies (Deb et al., 2002). The optimal solution gradually converges and stabilizes, and finally the Pareto front and these solution visualization results are generated. In the solution set, we picked the optimal solutions for each of the three objectives. Next, we applied the augmented achievement scalarizing function (AASF) decomposition method to derive an equal weight preferred solution, which we call the balanced solution for the three OFs (Singh and Deb, 2020). This evaluation involves comparing the solutions against predefined evaluation criteria and original DEM, ensuring that they align with the objectives of hydrological benefits and cost considerations.

## 3. Results

### 3.1. Pareto front solution set and optimizing solution efficiency

The TMMOO model was executed for 300 generations to optimize the three objectives subject to the constraints. The optimization process required approximately 20 minutes of computation to yield 300 generations (populations size = 200) in the Windows 10 environment (AMD Ryzen 7 5800h 3.20 GHz CPU, 16 GB memory).

Figure 6 shows the milestone results in the chosen 50th, 100th, 200th and 300th generations. the Pareto front is constantly approaching better spread when we examine any two of these objectives in pair-wise dimensions. The non-dominated solution numbers for the 50th, 100th, 200th and 300th generations are 55, 86, 197 and 200, respectively. Along with the step-by-step iteration, the solutions number increased and ultimately reached a convergence. The results of the 50th, 100th, and 200th generations have a clear iterative process of approaching the Pareto front boundary from a relatively scattered points pattern. By the 200th generation, its non-dominated solution number had approached the population size, and the scope exhibited a partly overlapping distribution with the final result, indicating the imminent completion of the convergence process.

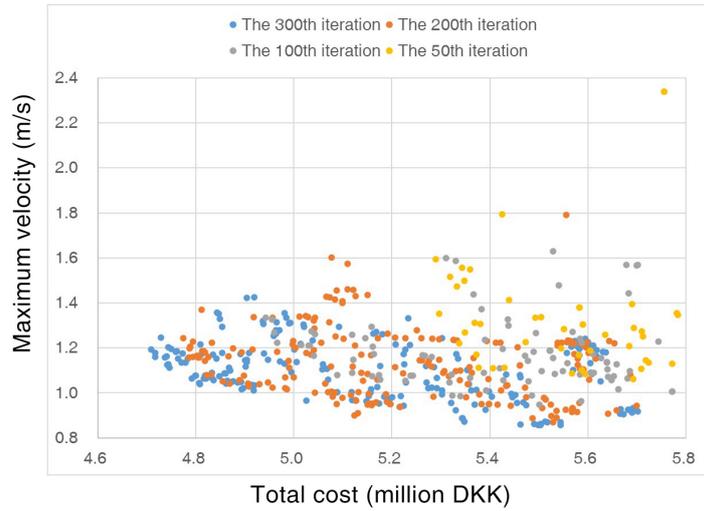

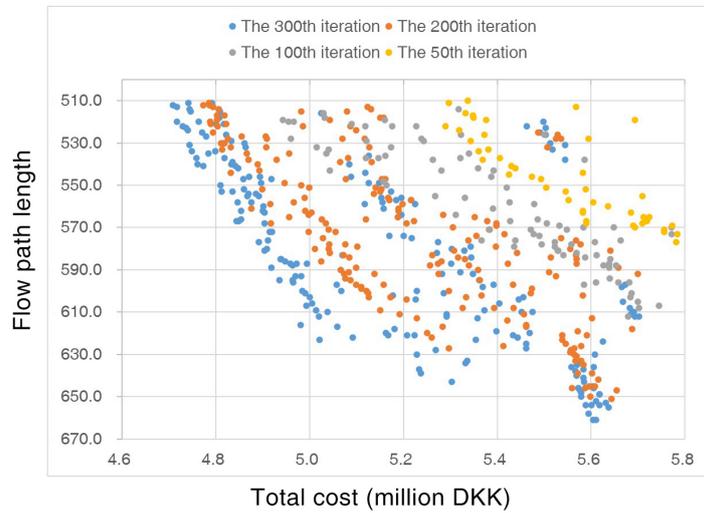

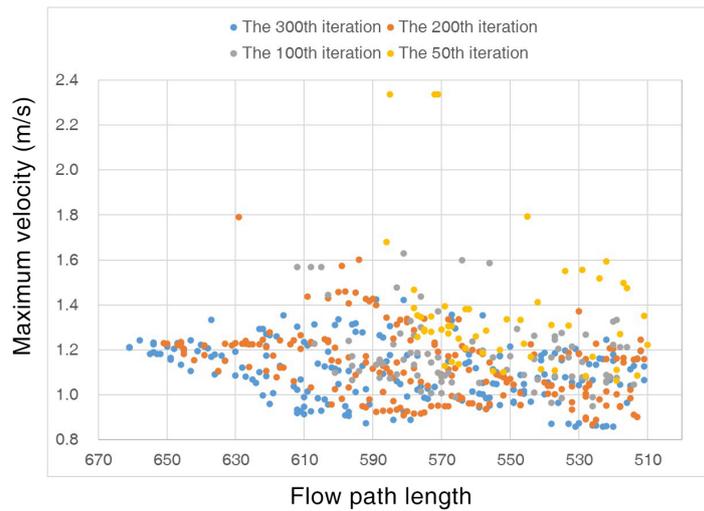

Figure 6 Pair-wise scatter plot of 50th, 100th, 200th and 300th generations in convergence process

Figure 7 illustrates the final Pareto front revealing the relationships among the three objectives. The total cost (million DKK) converged between 4.709 and 5.702. The flow path length converged between 511 and 661 with an original value is 510. The maximum runoff velocity (m/s) converged between 0.858 and 1.424, compared to the original value of 1.483. The pair-wise scatter plots indicate a negative correlation between the total cost objective and the flow path length objective, as well as the maximum velocity objective. This tendency indicates that longer flow path lengths and lower maximum velocities can be obtained as the total cost becomes more expensive. Between the flow path length objective and the maximum velocity objective, the solution set demonstrates a relatively uniform and dispersed distribution, indicating no obvious correlation.

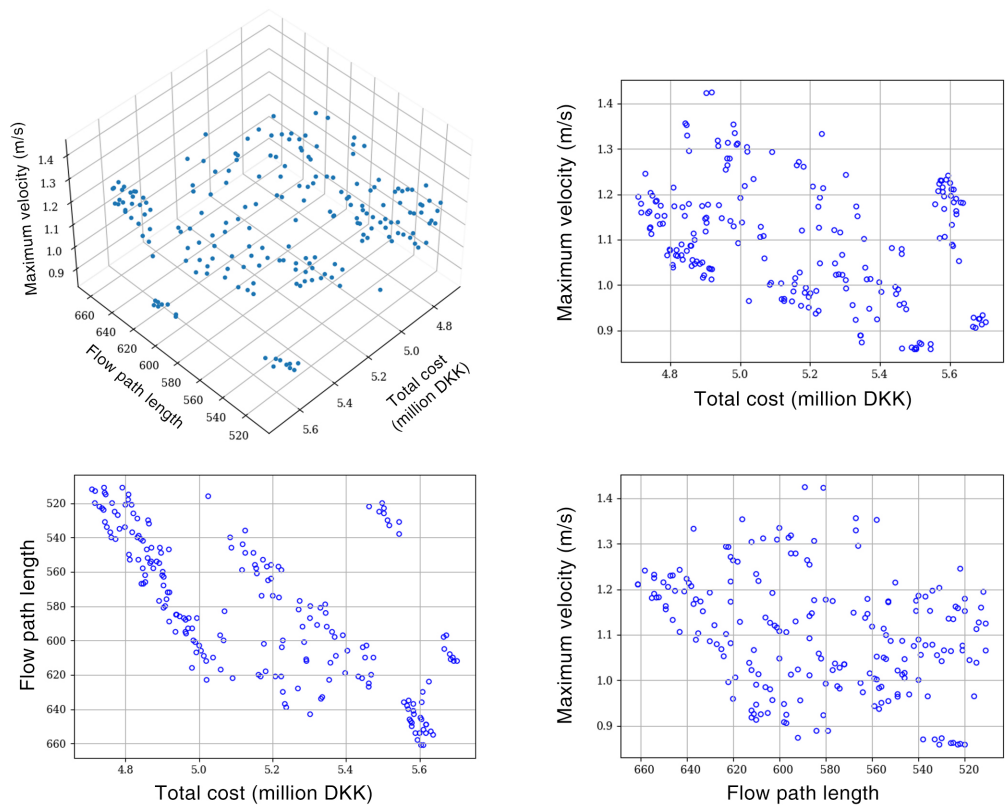

Figure 7 Pareto fronts presented by 3D scatter plot and pair-wise scatter plot of three objective functions

## 3.2. Specific preferred optimal solutions and decision-making support

After the verification of the simulation, the 200 solutions on the Pareto front can then be used to derive a practical solution when considering different qualitative requirements of different users. We took the equal weight preferred solution and optimal solutions for three single-objectives as example scenarios to validate the effectiveness of this model.

From the scenarios above, we can deduct that the equal weight preferred solution, refers to the balanced solution, has the most balanced hydrological-cost benefit with respect to all three OFs, which is the specific value shown in Figure 8. The other three solutions tend to be extremes, but they definitely reach the best scores with their preferred single objectives. In Figure 9, the minimum total cost solution is similar to the topography status quo, because less cost implies that there are not much terrain changes. The maximum flow path length solution can be observed as an increase in the length and aggregation of runoff paths in the central and southern areas, compared to the original site's more dispersed and single-branch runoff routes. The objective three preferred solution presents the minimum situation of the maximum velocity index.

The differences in the results of the target OFs are obvious and can be informed by examining the values of each solution. However, in a DEM with more than 3000 grid cells, the differences in visualization require careful comparisons to be discovered. We selected twenty solutions at every interval of ten solutions out of all 200 solutions to display diverse potential plans. Further statistical results and visualization were conducted on these Pareto Front points, as shown in supplementary materials (Appendix A).

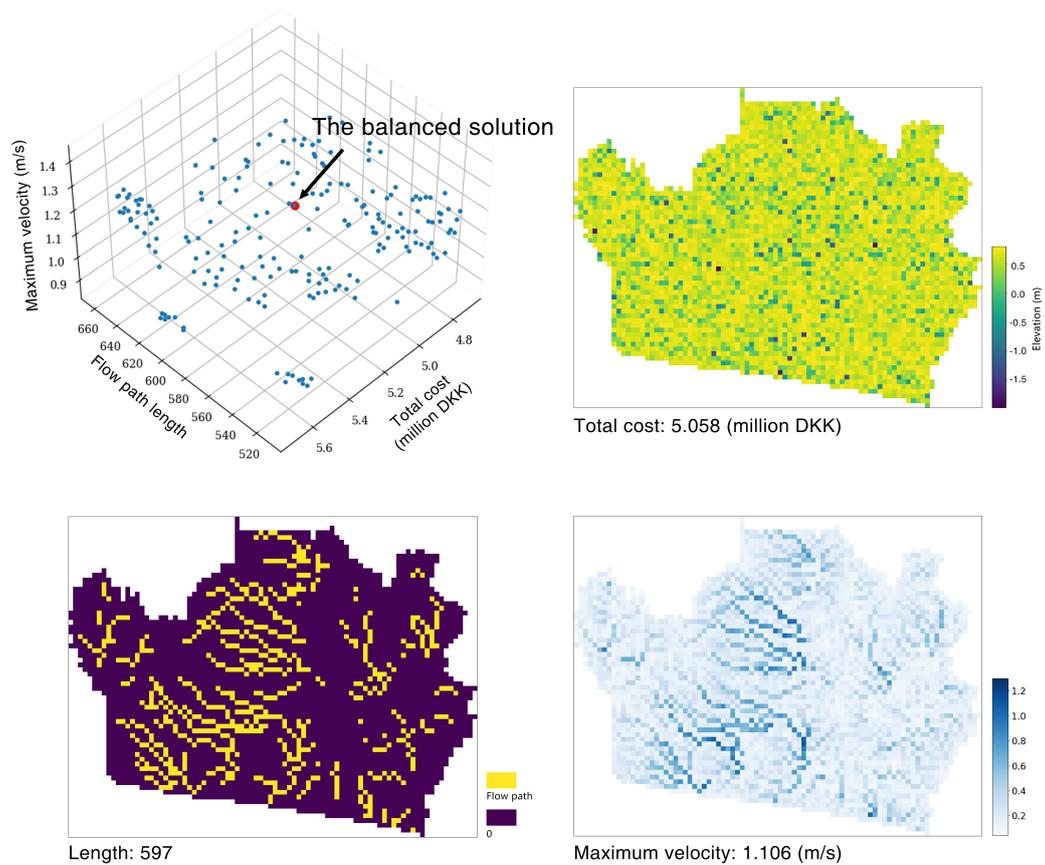

Figure 8 The balanced solution

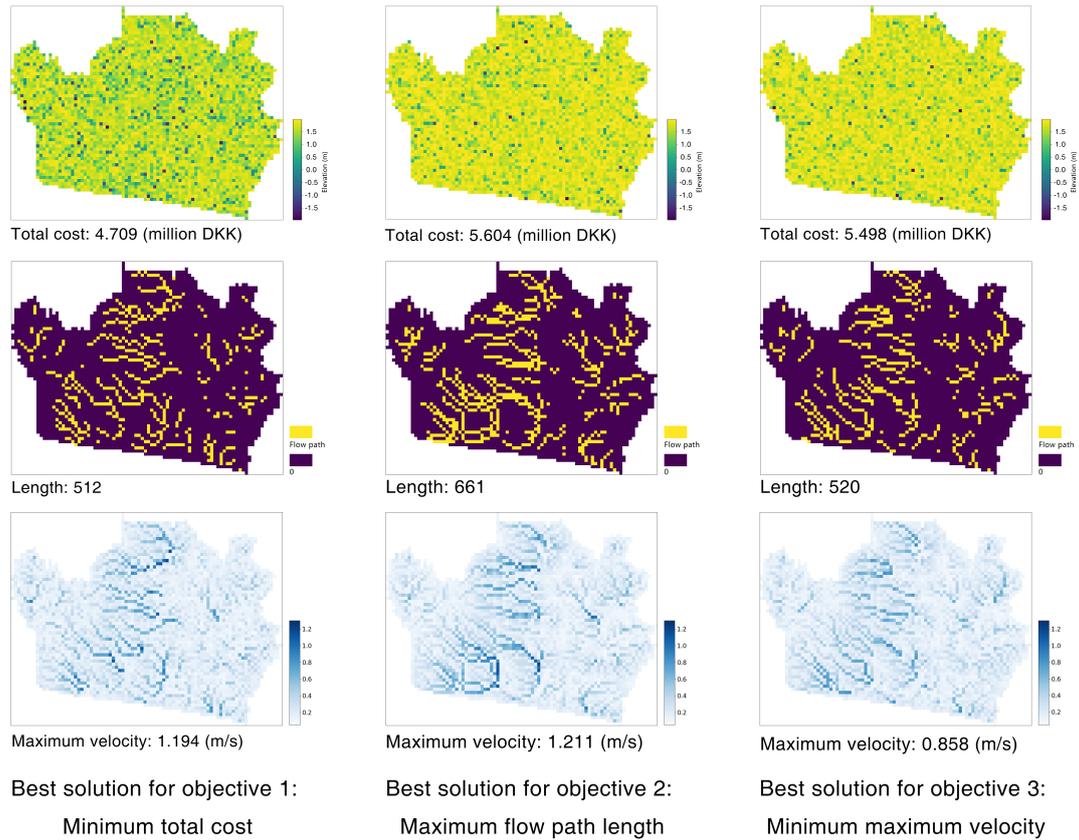

Figure 9 Optimal solutions for three single-objectives

## 4. Discussion

### 4.1. Findings and implication

Worldwide, investment and learning in stormwater management have increased substantially in the last decade (Chang et al., 2018; Liu et al., 2019; Wong, 2006). This increase indicates a willingness to pursue multi-functional, nature-based solutions in urban settings (Hobbie and Grimm, 2020). Meanwhile, the land use change and surface alternation resulting from urbanization yield a pessimistic situation in urban flood risk. Based on the results of our study, we advise stakeholders to review their workflow on how site terrain is designed, and to balance future terrain modification with hydrological and cost factors. Our framework outlines a straightforward approach to optimize the spatial distribution and change volume of earthwork, as well as evaluates how well these multi-objectives are met. Patterns of typical solutions such as optimal flow path length solution, optimal maximum velocity solution, optimal modification cost solution or balanced solution would support better decision making about how to deploy terrain modification works.

Our study emphasizes that the NSGA-II-based optimization method offers an efficient and highly customized solution to address uncertainties and repetition involving experience-based terrain modification. Especially for designers who are familiar with DEM data and analysis, this is convenient and user-friendly during the preliminary phase without the mass requirement of data and operation conditions. Also, this framework might inspire other DEM-based optimization research questions related to terrain modification. For example, flood risk assessment, ecosystem service evaluation, land use configuration or construction site selection are usually supported by GIS analysis and topography also plays a vital role in problem-solving processes (Cao et al., 2011; Miller et al., 2023; Xing et al., 2022). Such an integrated approach would be useful for informing stakeholders and coming to decisions on terrain planning and design enhancing the feasibility and practicality of stormwater management strategies.

## 4.2. Study limitation

To ensure practical support for stormwater management implementation and optimization framework replicability, it is important to acknowledge this study's limitations and how they may have affected the results. One of the limitations is that the confined number of variables for TMMOO has resulted in a restricted selection of DEM resolution and optimization precision. Furthermore, it would be a more advanced study if there were more comprehensive constraints to rationalize the position, earthwork aggregation and spatial layout of terrain modification, which might strengthen the practical operability of this optimization work. For example, identifying areas where terrain modification is feasible through surface classification to convert specific areas to variables for terrain modification. Also, combining terrain modifications in the form of terrain features (e.g., area limitation, length, width, height) of common stormwater management facilities (e.g., bio-retention ponds, swales, sunken green spaces) whose hydrologic benefits have been generally validated in cases and studies (Chui et al., 2016; Cutter and Pusch, 2021; Eckart et al., 2017; Pour et al., 2020). However, due to the Pymoo library's default setting, the limitations imposed on the customization coding ability achieved for this study limited our optimization framework.

In addition, terrain modification relies on complex mechanical operations and manual construction in practice. Due to the lack of application, the cost estimation of terrain modification is an ideally simulated solution in this study. It can be more convincing if the cost calculation can be further refined and compared with the actual project budgeted cost or the settlement amount. Last, it is meaningful to figure out how the outcomes of this study can be inserted into routine urban design references and ultimately into site guidelines for actual construction.

## 5. Conclusions

In this paper, a novel multi-objective optimization framework for terrain modification, TMMOO, is developed based on DEM data and related hydro-morphometric analysis and cost calculation.  The main conclusions are as follows:

1) Applying NSGA to solve the terrain modification optimization problem using DEM grid cells as variables has proven feasible. TMMOO demonstrates high efficiency and accuracy in generating solutions for three comprehensive objectives: maximizing flow path length, minimizing maximum runoff velocity, and minimizing construction cost.

2) Through comparison analysis and visualization of the obtained Pareto front solutions, non-dominated solutions display notable improvement. At different investment levels, the flow path lengths increased within 0.20% - 29.61% by connecting, widening, and extending the original flow path pattern. Maximum velocities within the study area have decreased within 3.98% - 42.14%, indicating possibly decreased erosion and potential damage to the site.

3) The versatility of TMMOO framework allows for potential expansion by incorporating additional objective functions and more comprehensive cost calculation. Various analysis functions based on DEM raster data and land cover attributes (e.g., land-use, surface type) can be integrated as constraints or objectives to further enhance terrain modifications' comprehensiveness and provide precise guidance for stormwater management planning.

It is important to note that TMMOO's solution sets are generated under several idealized conditions, which necessitates further refinement to align with practical stormwater management, landscape planning and construction considerations. The future integration of TMMOO and other hydrological or hydraulic models also holds better promise for guiding blue-green infrastructure or facilities layout planning and design. In conclusion, the TMMOO framework represents a small novel step forward in generated terrain modification optimization. As we continue to refine and adapt its procedures, we hope that TMMOO will play a pivotal role in real-world engineering guidance and contribute to effective stormwater management establishment ahead.

## Authorship contribution statement

**Hanwen Xu**: Conceptualization, Methodology, Optimization framework construction, Data analysis, Writing original draft, Review & editing, Visualization. **Mark Randall**: Conceptualization, Methodology, Review & editing, Supervision. **Lei Li**: Methodology, Review & editing, Coding assistance. **Yuyi Tan**: Review & editing, Coding assistance. **Thomas Balstrøm:** Methodology, Data analysis, Review & editing. All authors have read and agreed to the published version of the manuscript.

## Declaration of Competing Interest

The authors declare that they have no known competing financial interests or personal relationships that could have appeared to influence the work reported in this paper.

## Acknowledgment

The authors acknowledge China Scholarship Council (CSC), for providing funding for the research. Grant ID: 202106090018. Whitebox tools was downloaded from https://jblindsay.github.io/ghrg/WhiteboxTools/download.html, and there is also a front-end package available in R and Python to run it.

# References:

Ariza-Villaverde, A.B., Jiménez-Hornero, F.J., Gutiérrez De Ravé, E., 2013. Multifractal analysis applied to the study of the accuracy of DEM-based stream derivation. Geomorphology 197, 85-95.

Ariza-Villaverde, A.B., Jiménez-Hornero, F.J., Gutiérrez De Ravé, E., 2015. Influence of DEM resolution on drainage network extraction: A multifractal analysis. Geomorphology 241, 243-254.

Balstrøm, T., Crawford, D., 2018. Arc-Malstrøm: A 1D hydrologic screening method for stormwater assessments based on geometric networks. Comput. Geosci.-Uk 116, 64-73.

Blank, J., Deb, K., 2020. Pymoo: Multi-Objective Optimization in Python. IEEE Access 8, 89497-89509.

Burns, M.J., Fletcher, T.D., Walsh, C.J., Ladson, A.R., Hatt, B.E., 2012. Hydrologic shortcomings of conventional urban stormwater management and opportunities for reform. Landscape Urban Plan. 105(3), 230-240.

Cao, K., Batty, M., Huang, B., Liu, Y., Yu, L., Chen, J., 2011. Spatial multi-objective land use optimization: extensions to the non-dominated sorting genetic algorithm-II. International journal of geographical information science : IJGIS 25(12), 1949-1969.

Chang, N., Lu, J., Chui, T.F.M., Hartshorn, N., 2018. Global policy analysis of low impact development for stormwater management in urban regions. Land Use Policy 70, 368-383.

Chen, L., Dai, Y., Zhi, X., Xie, H., Shen, Z., 2018. Quantifying nonpoint source emissions and their water quality responses in a complex catchment: A case study of a typical urban-rural mixed catchment. J. Hydrol. 559, 110-121.

Chen, V., Bonilla Brenes, J.R., Chapa, F., Hack, J., 2021. Development and modelling of realistic retrofitted Nature-based Solution scenarios to reduce flood occurrence at the catchment scale. Ambio 50(8SI), 1462-1476.

Chen, Y., Samuelson, H.W., Tong, Z., 2016. Integrated design workflow and a new tool for urban rainwater management. J. Environ. Manage. 180, 45-51.

Chui, T.F.M., Liu, X., Zhan, W., 2016. Assessing cost-effectiveness of specific LID practice designs in response to large storm events. J. Hydrol. 533, 353-364.

Cutter, W.B., Pusch, A., 2021. The role of cost, scale, and property attributes in landowner choice of stormwater management option. Landscape Urban Plan. 209, 104040.

Dávila-Hernández, S., González-Trinidad, J., Júnez-Ferreira, H.E., Bautista-Capetillo, C.F., Morales De Ávila, H., Cázares Escareño, J., Ortiz-Letechipia, J., Robles Rovelo, C.O., López-Baltazar, E.A., 2022. Effects of the Digital Elevation Model and Hydrological Processing Algorithms on the Geomorphological Parameterization. Water-Sui. 14(15), 2363.

Deb, K., Jain, H., 2014. An Evolutionary Many-Objective Optimization Algorithm Using Reference-Point-Based Nondominated Sorting Approach, Part I: Solving Problems With Box Constraints. Ieee T. Evolut. Comput. 18(4), 577-601.

Deb, K., Pratap, A., Agarwal, S., Meyarivan, T., 2002. A fast and elitist multiobjective genetic algorithm: NSGA-II. Ieee T. Evolut. Comput. 6(PII S 1089-778X(02)04101-22), 182-197.

Eckart, K., McPhee, Z., Bolisetti, T., 2017. Performance and implementation of low impact development – A review. Sci. Total Environ. 607-608, 413-432.

Eckart, K., McPhee, Z., Bolisetti, T., 2018. Multiobjective optimization of low impact development stormwater controls. J. Hydrol. 562, 564-576.

Fletcher, T.D., Andrieu, H., Hamel, P., 2013. Understanding, management and modelling of urban hydrology and its consequences for receiving waters: A state of the art. Adv. Water Resour. 51, 261-

279.

Hobbie, S.E., Grimm, N.B., 2020. hobbie-grimm-2020-nature-based-approaches-to-managing-climate-change-impacts-in-cities. Philos. T. R. Soc. B 375, 20190124.

Johnson, D., Geisendorf, S., 2019. Are Neighborhood-level SUDS Worth it? An Assessment of the Economic Value of Sustainable Urban Drainage System Scenarios Using Cost-Benefit Analyses. Ecol. Econ. 158, 194-205.

Lee, J.G., Selvakumar, A., Alvi, K., Riverson, J., Zhen, J.X., Shoemaker, L., Lai, F., 2012. A watershed-scale design optimization model for stormwater best management practices. Environ. Modell. Softw. 37, 6-18.

Lee, J.H., Bang, K.W., 2000. Characterization of urban stormwater runoff. Water Res. 34(6), 1773-1780.

Leng, L., Jia, H., Chen, A.S., Zhu, D.Z., Xu, T., Yu, S., 2021. Multi-objective optimization for green-grey infrastructures in response to external uncertainties. Sci. Total Environ. 775, 145831.

Liu, L., Fryd, O., Zhang, S., 2019. Blue-Green Infrastructure for Sustainable Urban Stormwater Management—Lessons from Six Municipality-Led Pilot Projects in Beijing and Copenhagen. Water (Basel) 11(10), 2024.

Liu, Z., Han, Z., Shi, X., Liao, X., Leng, L., Jia, H., 2023b. Multi-objective optimization methodology for green-gray coupled runoff control infrastructure adapting spatial heterogeneity of natural endowment and urban development. Water Res. 233, 119759.

Melesse, A.M., Graham, W.D., 2004. STORM RUNOFF PREDICTION BASED ON A SPATIALLY DISTRIBUTED TRAVEL TIME METHOD UTILIZING REMOTE SENSING AND GIS1. J. Am. Water Resour. As. 40(4), 863-879.

Miller, J.D., Vesuviano, G., Wallbank, J.R., Fletcher, D.H., Jones, L., 2023. Hydrological assessment of urban Nature-Based Solutions for urban planning using Ecosystem Service toolkit applications. Landscape Urban Plan. 234, 104737.

Miltner, R.J., White, D., Yoder, C., 2004. The biotic integrity of streams in urban and suburbanizing landscapes. Landscape Urban Plan. 69(1), 87-100.

Nesshöver, C., Assmuth, T., Irvine, K.N., Rusch, G.M., Waylen, K.A., Delbaere, B., Haase, D., Jones-Walters, L., Keune, H., Kovacs, E., Krauze, K., Külvik, M., Rey, F., van Dijk, J., Vistad, O.I., Wilkinson, M.E., Wittmer, H., 2017. The science, policy and practice of nature-based solutions: An interdisciplinary perspective. Sci. Total Environ. 579, 1215-1227.

Nicola, F., Jorg, B., J. Martin, H., Rudolph, A., 1999. Changing soil and surface conditions during rainfall Single rainstormrsubsequent rainstorms. Catena 37, 355-375.

O'Callaghan, J.F., Mark, D.M., 1984. The Extraction of Drainage Networks from Digital Elevation Data. COMPUTER VISION GRAPHICS AND IMAGE PROCESSING 28(3), 323-344.

Pour, S.H., Wahab, A.K.A., Shahid, S., Asaduzzaman, M., Dewan, A., 2020. Low impact development techniques to mitigate the impacts of climate-change-induced urban floods: Current trends, issues and challenges. Sustainable Cities and Society 62, 102373.

Randall, M., Støvring, J., Henrichs, M., Bergen Jensen, M., 2020. Comparison of SWMM evaporation and discharge to in-field observations from lined permeable pavements. Urban Water J. 17(6), 491-502.

Saadatpour, M., Delkhosh, F., Afshar, A., Solis, S.S., 2020. Developing a simulation-optimization approach to allocate low impact development practices for managing hydrological alterations in urban watershed. Sustainable Cities and Society 61, 102334.

Salvan, L., Abily, M., Gourbesville, P., Schoorens, J., 2016a. Drainage System and Detailed Urban

Topography: Towards Operational 1D-2D Modelling for Stormwater Management. Procedia Engineering 154, 890-897.

Shishegar, S., Duchesne, S., Pelletier, G., 2018. Optimization methods applied to stormwater management problems: a review. Urban Water J. 15(3), 276-286.

Sidek, L.M., Chua, L.H.C., Azizi, A.S.M., Basri, H., Jaafar, A.S., Moon, W.C., 2021. Application of PCSWMM for the 1-D and 1-D-2-D Modeling of Urban Flooding in Damansara Catchment, Malaysia. APPLIED SCIENCES-BASEL 11(930019).

Singh, H.K., Deb, K., 2020. Investigating the equivalence between PBI and AASF scalarization for multi-objective optimization. Swarm and Evolutionary Computation 53, 100630.

Sun, H., Dong, Y., Lai, Y., Li, X., Ge, X., Lin, C., 2022. The Multi-Objective Optimization of Low-Impact Development Facilities in Shallow Mountainous Areas Using Genetic Algorithms. Water-Sui. 14(19), 2986.

Tang, S., Jiang, J., Shamseldin, A.Y., Shi, H., Wang, X., Shang, F., Wang, S., Zheng, Y., 2022. Comprehensive Optimization Framework for Low Impact Development Facility Layout Design with Cost‑Benefit Analysis: A Case Study in Shenzhen City, China. ACS ES&T Water 2(1), 63-74.

Wang, M., Jiang, Z., Zhang, D., Zhang, Y., Liu, M., Rao, Q., Li, J., Keat Tan, S., 2023. Optimization of integrating life cycle cost and systematic resilience for grey-green stormwater infrastructure. Sustainable Cities and Society 90, 104379.

Wong, T.H.F., 2006. Water sensitive urban design - the journey thus far. Australian journal of water resources 10(3), 213-222.

Xie, M., Cheng, Y., Dong, Z., 2022. Study on Multi-Objective Optimization of Sponge Facilities Combination at Urban Block Level: A Residential Complex Case Study in Nanjing, China. Water-Sui. 14(20), 3292.

Xing, Y., Chen, H., Liang, Q., Ma, X., 2022. Improving the performance of city-scale hydrodynamic flood modelling through a GIS-based DEM correction method. Nat. Hazards 112(3), 2313-2335.

Xu, H., Randall, M., Fryd, O., 2023. Urban stormwater management at the meso-level: A review of trends, challenges and approaches. J. Environ. Manage. 331, 117255.

Xu, T., Engel, B.A., Shi, X., Leng, L., Jia, H., Yu, S.L., Liu, Y., 2018. Marginal-cost-based greedy strategy (MCGS): Fast and reliable optimization of low impact development (LID) layout. Sci. Total Environ. 640-641, 570-580.

Xu, T., Jia, H., Wang, Z., Mao, X., Xu, C., 2017. SWMM-based methodology for block-scale LID-BMPs planning based on site-scale multi-objective optimization: a case study in Tianjin. Front. Env. Sci. Eng. 11(4).

Yang, B., Zhang, T., Li, J., Feng, P., Miao, Y., 2023. Optimal designs of LID based on LID experiments and SWMM for a small-scale community in Tianjin, north China. J. Environ. Manage. 334, 117442.

Yang, W., Zheng, C., Jiang, X., Wang, H., Lian, J., Hu, D., Zheng, A., 2023. Study on urban flood simulation based on a novel model of SWTM coupling D8 flow direction and backflow effect. J. Hydrol. 621, 129608.

Yao, Y., Li, J., Lv, P., Li, N., Jiang, C., 2022. Optimizing the layout of coupled grey-green stormwater infrastructure with multi-objective oriented decision making. J. Clean. Prod. 367, 133061.

Yin, D., Chen, Y., Jia, H., Wang, Q., Chen, Z., Xu, C., Li, Q., Wang, W., Yang, Y., Fu, G., Chen, A.S., 2021. Sponge city practice in China: A review of construction, assessment, operational and maintenance. J. Clean. Prod. 280, 124963.

Zhang, K., Chui, T.F.M., 2018. A comprehensive review of spatial allocation of LID-BMP-GI practices:

Strategies and optimization tools. Sci. Total Environ. 621, 915-929.